\documentclass[preprintnumbers,floatfix,12pt,superscriptaddress,nofootinbib]{revtex4}

\usepackage{amsfonts}
\usepackage{amssymb,epsfig}
\usepackage{latexsym}
\usepackage{amsmath}
\usepackage{graphicx}

\begin{document}

\title{ Statefinder diagnosis and the interacting ghost model of dark energy}
\author{M. Malekjani$^{1}$ \footnote{%
Email: \text{malekjani@basu.ac.ir}},A. Khodam-Mohammadi \footnote{%
Email: \text{khodam@basu.ac.ir}}}
\affiliation{Department of Physics,
Faculty of Science, Bu-Ali Sina University, Hamedan 65178, Iran\\}

\begin{abstract}
\vspace*{1.5cm} \centerline{\bf Abstract} \vspace*{1cm} A new model
of dark energy namely "ghost dark energy model" has recently been
suggested to interpret the positive acceleration of cosmic
expansion. The energy density of ghost dark energy is proportional
to the hubble parameter. In this paper we perform the statefinder
diagnostic tool for this model both in flat and non-flat universe.
We discuss the dependency of the evolutionary trajectories in $s-r$
and $q-r$ planes on the interaction parameter between dark matter
and dark energy as well as the spatial curvature parameter of the
universe. Eventually, in the light of SNe+BAO+OHD+CMB observational
data, we plot the evolutionary trajectories in $s-r$ and $q-r$
planes for the best fit values of the cosmological parameters and
compare the interacting ghost model with other dynamical dark energy
models. We show that the evolutionary trajectory of ghost dark
energy in statefinder diagram is similar to holographic dark energy
model. It has been shown that the statefinder location of
$\Lambda$CDM is in good agreement with observation and therefore the
dark energy models whose current statefinder values are far from the
$\Lambda$CDM point can be ruled out.

\end{abstract}
\maketitle

\newpage
\section{Introduction}
Nowadays it is strongly believed that our universe expands under an
accelerated expansion. The various cosmological data gathered from
SNe Ia \cite{c1}, WMAP \cite{c2}, SDSS \cite{c3} and X-ray \cite{c4}
experiments have provided the main evidences for this cosmic
acceleration. Within the framework of standard cosmology, a dark
energy component with negative pressure is responsible for this
acceleration. Up to now many theoretical models have been proposed
to interpret the behavior of dark energy. The first and simple
candidate is the Einstein's cosmological constant with the time -
independent equation of state $w_{\Lambda}=-1$. The cosmological
constant suffers from tow deep theoretical problems namely the
"fine-tuning" and "cosmic coincidence". In addition to cosmological
constant, dynamical dark energy model with time- varying equation of
state have been investigated to interpret the cosmic acceleration.
The scalar field models such as quintessence \cite{c5}, phantom
\cite{c6}, quintom \cite{c7}, K-essence \cite{c8}, tachyon \cite{c9}
and dilaton \cite{c10} together with interacting dark energy models
such as holographic \cite{c11} and agegraphic \cite{c12} models are
the examples of dynamical dark energy models. The interacting dark
energy models have been constructed within the framework of quantum
gravity, by introducing the new degree of
freedom or by modifying the theory of gravity \cite{c13,c14,c144}.\\
 Recently, the
Veneziano ghost dark energy has been attracted a deal of attention
in the dynamical DE category. The Veneziano ghost is proposed to
solve the $U(1)$ problem in low-energy effective theory of QCD
\cite{witten} and has no contribution in the flat Minkowski
spacetime. In curved spacetime, however, it makes a small energy
density proportional to $\Lambda^3_{QCD}H$, where $\Lambda_{QCD}$ is
QCD mass scale and $H$ is Hubble parameter. This small vacuum energy
density can be considered as a driver engine for evolution of the
universe. It is worthwhile to mention that this model is totally
arisen from standard model and general relativity. Therefore one
needs not to introduce any new parameter or new degree of freedom
and this fact is the most advantages of ghost DE. With
$\Lambda_{QCD}\sim 100Mev$ and $H\sim 10^{-33}ev$, the right order
of observed DE density can be given by ghost DE. This numerical
coincidence also shows that this model gets ride of fine tuning
problem \cite{urban,ohta} Many authors have already suggested DE
model with energy density as $\rho= \alpha H$ \cite{bjorken}.\\
Recent observational data gathered from the Abell Cluster A586
support the interaction between dark matter and dark energy
\cite{c15}. However the strength of this interaction is not clearly
identified \cite{c16}. \\
Since many theoretical dark energy models have been proposed to
explain the accelerated expansion of the universe, therefore the
sensitive test which can differentiate between these models is
required. The Hubble parameter, $H=\dot{a}/a$, (first time
derivative) and the deceleration parameter $q=-\ddot{a}H^2/a$
(second time derivative) are the geometrical parameters to describe
the expansion history of the universe. $\dot{a}>0$ or $H>0$ means
the expansion of the universe. Also $\ddot{a}>0$, i.e. $q<0$,
indicates the accelerated expansion of the universe. Since the
various dark energy models give $H>0$, $q<0$ at the percent time,
the Hubble and deceleration parameters can not discriminate dark
energy models. For this aim we need a higher order of time
derivative of scale factor. Sahni et al. \cite{sah03} and Alam et
al. \cite{alamb}, by using the third time derivative of scale
factor, introduced the statefinder pair \{s,r\} in order to remove
the degeneracy of $H$ and $q$ at the present time. The statefinder
pair is given by
\begin{equation}\label{state1}
r=\frac{\dddot{a}}{aH^3},s=\frac{r-1}{3(q-1/2)}
\end{equation}
Depending the statefinder diagnostic tool on the scale factor
indicates that the statefinder parameters are geometrical. The scale
factor $a(t)$ can be expanded near the present time $t_0$ as follows
\begin{equation}
a(t)=1+H_0(t-t_0)-\frac{1}{2}q_0H_0^2(t-t_0)^2+\frac{1}{6}r_0H_0^3(t-t_0)^3+...
\end{equation}
where we consider $a(t_0)=1$ and $H_0$, $q_0$, $r_0$ are the present
values of the Hubble parameter, deceleration parameter and former
statefinder parameter, respectively. Up to now, the various dark
energy models have been studied from the viewpoint of statefinder
diagnostic. These models have different evolutionary trajectories in
\{s, r\} plane, therefore the statefinder tool can discriminate
these models. The well known $\Lambda$CDM model is related to the
fixed point \{s=0,r=1\} in the $s-r$ plane \cite{sah03}. The
distance of the current value of statefinder pair $\{s_0, r_0\}$ for
a given dark energy model from the fixed point \{s=0,r=1\} is a
valuable criterion to a model. In addition, the distance of current
statefinder values of a given dark energy model from the constrained
observational value $\{s_0, r_0\}$ is a good tool to test a model.\\
The dynamical dark energy models that have been investigated by
statefinder
diagnostic tool are:\\
the quintessence DE model \cite{sah03, alamb} , the interacting
quintessence models \cite{zim, zhang055}, the holographic dark
energy models \cite{zhang056, zhang057} , the holographic dark
energy model in non-flat universe \cite{setar}, the phantom model
\cite{chang}, the tachyon \cite{shao}, the generalized chaplygin gas
model \cite{malek11}, the interacting new agegraphic DE model in
flat and non-flat universe  \cite{ zhang058, malek12}, the
agegraphic dark energy model with and without interaction in flat
and non-flat universe \cite{wei077, malek13}, the new holographic
dark energy model \cite{malek20} and the interacting polytropic gas
model \cite{poly99}.\\
In this work we investigate the interacting ghost
dark energy model by statefinder diagnostic tool. The statefinder
can be applied to diagnose different cases of the model, including
different model parameters and different contributions of spatial
curvature. First, we perform the statefinder diagnostic in flat
universe in sect. II, then we generalize our work to the non flat
universe in sect. III. In sect.IV, the statefinder diagnostic has
been discussed based on recent observational data. This work is
concluded in sect. V.

\section{Interacting ghost dark energy model in flat universe}
Let us first consider the interacting ghost dark energy in the flat
Friedmann-Robertson-Walker (FRW) universe. The corresponding
Friedmann equation in this case is given by
\begin{equation}\label{fridt}
H^{2}=\frac{1}{3M_{p}^{2}}(\rho _{m}+\rho _{\Lambda})
\end{equation}%
where $H$ and $M_p$ are the Hubble parameter and the reduced Planck
mass, respectively.\\
The energy density of ghost dark energy is given by \cite{ghost1}
\begin{equation}\label{statet}
\rho_{\Lambda}=\alpha H
\end{equation}
where $\alpha$ is a constant of the model. The dimensionless energy
densities are defined as
\begin{equation}\label{denergyt}
\Omega_{m}=\frac{\rho_m}{\rho_c}=\frac{\rho_m}{3M_p^2H^2}, ~~~\\
\Omega_{\Lambda}=\frac{\rho_{\Lambda}}{\rho_c}=\frac{\rho_{\Lambda}}{3M_p^2H^2}~~\\
\end{equation}
Using (\ref{denergyt}), the Friedmann equation (\ref{fridt}) can be
written as
\begin{equation}
\Omega _{m}+\Omega _{\Lambda}=1.  \label{Freqt}
\end{equation}%
 In a universe dominated by interacting dark energy and
dark matter, the total energy density, $\rho=\rho_m+\rho_{\Lambda}$,
satisfies the following conservation equation
\begin{equation}
\dot{\rho}+3H(\rho+p)=0
\end{equation}
However, by considering the interaction between dark energy and dark
matter, the energy density of dark energy and dark matter does not
conserve separately and the conservation equation for each component
is given by
\begin{eqnarray}
\dot{\rho _{m}}+3H\rho _{m}=Q, \label{contmt}\\
\dot{\rho _{\Lambda}}+3H(\rho_{\Lambda}+p_{\Lambda})=-Q,
\label{contdt}
\end{eqnarray}%
where  $Q$ represents the interaction between dark matter and dark
energy. It is worth noting that in equation (\ref{contmt}) the right
hand side of (\ref{contmt}), same as left hand side, should be as a
function of inverse of time. The simple choice is that the
interaction quantity $Q$ can be considered as a function of Hubble
parameter $H$ such as one of the following forms: (i) $Q\propto
H\rho_{\Lambda}$, (ii) $Q\propto H\rho_{m}$ and (iii) $Q\propto
H(\rho_{m}+\rho_{\Lambda})$. One can assume the above three forms as
$Q=\Gamma \rho_{\Lambda}$, where for case (i) $\Gamma=3b^2H$, for
case (ii) $\Gamma=3b^2H \frac{\Omega_m}{\Omega_{\Lambda}}$ and for
case (iii) $\Gamma=3b^2H \frac{1}{\Omega_{\Lambda}}$. The parameter
$b$ is a coupling constant indicating the strength of interaction
between dark matter and dark energy \cite{c21}.
 The interaction between dark energy and dark matter is also studied in
\cite{c233}. Here we assume the third form of interaction for $Q$.\\
Taking the time derivative from both side of Friedmann equation
(\ref{fridt}) and using (\ref{Freqt}, \ref{contmt}, \ref{contdt}) as
well as the relation $p_{\Lambda}=w_{\Lambda}\rho_{\Lambda}$, one
can obtain
\begin{equation}\label{hdott}
\frac{\dot{H}}{H^2}=-\frac{3}{2}[1+w_{\Lambda}\Omega_{\Lambda}]
\end{equation}
Inserting the third form of interaction term
$Q=\Gamma\rho_{\Lambda}=3b^2H \frac{1}{\Omega_{\Lambda}}
\rho_{\Lambda}$ in the right hand side of (\ref{contdt}) and using
the relations (\ref{statet}), (\ref{hdott}), the equation of state
for interacting ghost dark energy in the flat universe can be
obtained as
\begin{equation}\label{statet1}
w_{\Lambda}=\frac{-1}{2-\Omega_{\Lambda}}(1+\frac{2b^2}{\Omega_{\Lambda}})
\end{equation}
In the limiting case of non-interacting flat universe (i.e., $b=0$
and $\Omega_k=0$), Eq.(\ref{statet1}) reduces to
\begin{equation}
w_{\Lambda}=-\frac{1}{2-\Omega_{\Lambda}}
\end{equation}
which is in agreement with \cite{shaikh9}. At the early time when
$\Omega_{\Lambda}<<1$, we can see $w_{\Lambda}=-1/2$ and at the late
time when $\Omega_{\Lambda}\sim1$, one can see $w_{\Lambda}=-1$.
Therefore the ghost dark energy mimics the cosmological constant at
the late time. The evolution of EoS parameter of ghost model has
been studied in \cite{shaikh9}. It has been shown that the
interacting ghost dark energy model can cross the phantom divide for
$b^2>0.1$.\\
Using (\ref{hdott}), the deceleration parameter $q$ in this model
can be obtained as
\begin{equation}\label{decet1}
q=-1-\frac{\dot{H}}{H^2}=\frac{1}{2}+\frac{3}{2}w_{\Lambda}\Omega_{\Lambda}
\end{equation}
It is clear that at the early time ( when
$\Omega_{\Lambda}\rightarrow 0$) we have $q=1/2$ which is equal to
the value of deceleration parameter obtained in CDM model. Therefore
in ghost model, the decelerated expansion phase ($q>0$) at the early
time can be achieved.
 At the late time ( when
$\Omega_{\Lambda}\sim1$ and $w_{\Lambda}=-1$), we see that $q=-1$,
which represents the accelerated expansion
($q<0$) in dark energy dominated universe, as expected.\\
Tacking the time derivative of dark energy density parameter in
(\ref{denergyt}) and using the ghost dark energy density
(\ref{statet}), we have
\begin{equation}\label{statet99}
\dot{\Omega_{\Lambda}}=-\frac{\alpha\dot{H}}{3M_p^2H^2}
\end{equation}
Using (\ref{decet1}) and
$\dot{\Omega_{\Lambda}}=H\Omega_{\Lambda}^{\prime}$ yields
\begin{equation}\label{dif1}
\Omega_{\Lambda}^{\prime}=\frac{3}{2}\Omega_{\Lambda}(1+w_{\Lambda}\Omega_{\Lambda})
\end{equation}
where prime denotes the derivative with respect to $\ln{a}$.
 Tacking the time derivative of (\ref{hdott}) and using
(\ref{denergyt}), (\ref{contdt}) and (\ref{statet}) we obtain
\begin{eqnarray}\label{ddotht}
\frac{\ddot{H}}{H^3}=\frac{9}{4}w_{\Lambda}\Omega_{\Lambda}(w_{\Lambda}\Omega_{\Lambda}+3)-
\frac{3}{2}\Omega_{\Lambda}w_{\Lambda}^{\prime} +\frac{18}{4}
\end{eqnarray}
We now find the statefinder parameters $\{s, r\}$ for the
interacting ghost dark energy model in the flat universe.  From the
definition of $q$ and $H$, the parameter $r$ in (\ref{state1}) can
be written as
\begin{equation}\label{stateft}
r=\frac{\ddot{H}}{H^3}-3q-2.
\end{equation}
Substituting the relations (\ref{decet1}) and (\ref{ddotht}) in
(\ref{stateft}), the parameter $r$ is obtained as
\begin{equation}\label{stater1t}
r=1+\frac{9}{4}w_{\Lambda}\Omega_{\Lambda}(w_{\Lambda}\Omega_{\Lambda}+1)-
\frac{3}{2}\Omega_{\Lambda}w_{\Lambda}^{\prime}
\end{equation}
Inserting Eqs. (\ref{decet1}) and (\ref{stater1t}) in the parameter
$s$ of (\ref{state1}) obtains
\begin{equation}\label{st6}
s=\frac{1}{2}(1+w_{\Lambda}\Omega_{\Lambda})-\frac{w_{\Lambda}^{\prime}}{3w_{\Lambda}}
\end{equation}
At the late time ( when $\Omega_{\Lambda}\rightarrow 1$ ), by
inserting $w_{\Lambda}=-1$ and therefore $w_{\Lambda}^{\prime}=0$,
the relations (\ref{stater1t}) and (\ref{st6}) reduce to the
constant values ($r=1$, $s=0$) which refers the statefinder
parameters of standard $\Lambda$CDM model in the flat universe.
Therefore, from the viewpoint of statefinder diagnostic, the ghost
dark energy mimics the cosmological constant at the late time.\\
By numerical solving of Eqs. (\ref{stater1t}) and (\ref{st6}), we
obtain the evolutionary trajectory of interacting ghost dark energy
in the statefinder plane. It should be noted that in Eqs.
(\ref{stater1t}) and (\ref{st6}) the evolution of $w_{\Lambda}$ and
$\Omega_{\Lambda}$ are governed by Eqs. (\ref{dif1})
and(\ref{statet1}), respectively. In statefinder plane, the
horizontal axis is defined by the parameter $s$ and vertical axis by
the parameter $r$. In this diagram, the standard $\Lambda$CDM model
corresponds to the fixed point $\{r = 1, s = 0\}$.\\
In Fig.(1), we plot the evolutionary trajectories of ghost dark
energy model in the flat universe in $s-r$ plane for different
illustrative values of interaction parameter $b$. Here we adopt the
current values of cosmological parameters $\Omega_{\Lambda}$ and
$\Omega_m$ as $0.7$ and $0.3$, respectively. The standard $\Lambda
CDM$ fixed point $\{r=1,s=0\}$ is indicated by star symbol in this
diagram. The colored circles on the curves show the present values
of statefindr pair $\{s_0, r_0\}$. By expanding the universe, the
trajectories in $s-r$ plane start from right to left. The parameter
$r$ decreases, then increases to the constant value $r=1$ at the
late time. While the parameter $s$ deceases from the positive value
at the early time to the constant value $s=0$ at the late time.
Different values of interaction parameter $b$ result the different
evolutionary trajectories in $s-r$ plane. Hence the statefinder
analysis can discriminate the interacting ghost dark energy model
with different interaction parameter. For larger value of $b$, the
present values of $s_0$ and $r_0$ decreases. The distance of the
point $\{s_0, r_0\}$ form the $\Lambda CDM$ fixed point
$\{s=0,r=1\}$ becomes larger for larger values of interaction
parameter $b$. Fig.(1) also shows that the interacting ghost dark
energy model mimics the $\Lambda$CDM model at the late time. This
behavior of ghost dark energy is similar to the holographic
\cite{zhang056,zhang057,setar}, new agegraphic
\cite{zhang058,malek12}, chaplygin gas \cite{chap2}, generalized
chaplygin gas \cite{malek11} and yang- mils \cite{yang} models of
dark energy
 in which they also mimic the $\Lambda$CDM model at the late time. \\
Unlike the above models, the agegraphic dark energy model
\cite{wei077,malek13} and polytropic gas model \cite{poly99} mimic
the $\Lambda$CDM model at the early stage of the evolution of the
universe. The evolutionary trajectories of holographic dark energy
under granda-Oliveros IR cut-off (new holographic model)
\cite{malek20} and also tachyon dark energy model \cite{shao} in
$s-r$ plane pass through the $\Lambda$CDM fixed point at the middle
of the evolution of the universe. The other interesting note is that
the evolution of ghost dark energy model in $s-r$ plane is similar
to the evolution of holographic model of dark energy with the model
parameter $c=1$ in this plane (i.e., see Fig.(3) of \cite{zhang057}
and upper panel of Fig.(1) in \cite{setar}).\\
 Also, it is of interest to discuss the dynamical behavior of
ghost dark energy in $q-r$ plane. In $q-r$ plane, we use the
geometrical quantity $q$ instead of the parameter $s$ at the
horizontal axis. In Fig.(2), by solving Eqs.(\ref{decet1}) and
(\ref{stater1t}), the evolutionary trajectories of ghost dark energy
in flat universe is plotted for different values of interaction
parameter $b$ in $q-r$ plane. Same as statefinder analysis, the
$q-r$ analysis can discriminate different dark energy models. By
expanding the universe, the trajectories start from right to left.
The parameter $r$ decrease, then increases to the constant value
$r=1$ at the late time. While the parameter $q$ decreases from the
positive value ( indicating the decelerated expansion) at the early
time to the negative value (representing the accelerated expansion)
at the late time. Here we see the different evolutionary
trajectories for different interaction parameters $b$. The current
value $\{q_0, r_0\}$ can also be affected by interaction parameter.
Increasing the interaction parameter $b$ causes both the parameters
$r$ and $q$ becomes smaller.
\begin{center}
\begin{figure}[!htbp]
\includegraphics[width=10cm]{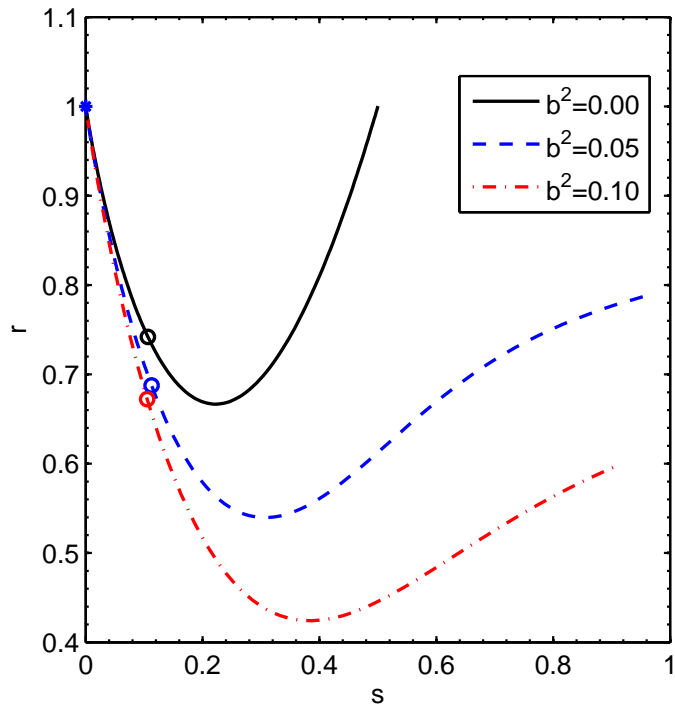}\\
~~~~~~ ~~~~~~\caption{ The evolutionary trajectories in $s-r$ plane
for interacting ghost dark energy model in the flat universe with
the cosmological parameters $\Omega_{m0}=0.3$ and $\Omega_{\Lambda
0}=0.7$. The location of standard $\Lambda$CDM fixed point is
indicated by star symbol. The colored circle points are the location
of present values of statefinder pair $\{s_0, r_0\}$ for different
interaction parameter as described in legend.}
 \end{figure}
 \end{center}

\begin{center}
\begin{figure}[!htbp]
\includegraphics[width=10cm]{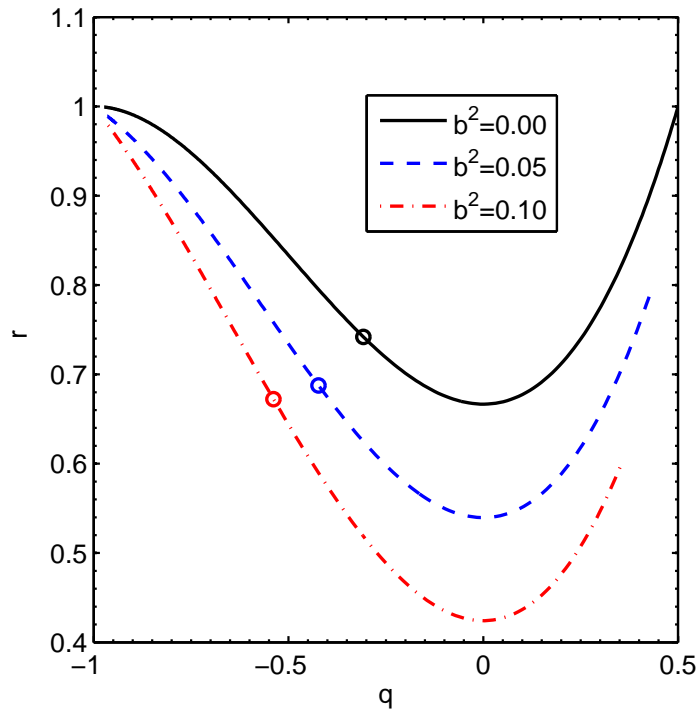}\\
~~~~~~ ~~~~~~\caption{ The evolutionary trajectories in $q-r$ plane
for interacting ghost dark energy model in the flat universe with
the cosmological parameters $\Omega_{m0}=0.3$ and $\Omega_{\Lambda
0}=0.7$. The colored circle points are the location of present
values of statefinder pair $\{q_0, r_0\}$ for different interaction
parameter as described in legend.}
 \end{figure}
 \end{center}

\section{Interacting ghost dark energy model in a non flat universe}
In this section we generalize our work in previous section to the
non flat universe. The Friedmann equation in this case is given by
\begin{equation}\label{frid1}
H^{2}+\frac{k}{a^{2}}=\frac{1}{3M_{p}^{2}}(\rho _{m}+\rho
_{\Lambda})
\end{equation}%
where $k=1,0,-1$ is a spatial curvature parameter corresponding to
the closed, flat and open universe, respectively. The dimensionless
energy densities of dark energy and dark matter have been defined in
(\ref{denergyt}) and dimensionless energy density corresponding to
the spatial curvature is given as $\Omega_k=\frac{k}{a^2H^2}$.
Therefore the Friedmann equation (\ref{frid1}) in terms of
dimensionless parameters is written as
\begin{equation}
\Omega _{m}+\Omega _{\Lambda}=1+\Omega _{k}.  \label{Freq2}
\end{equation}%
Same as previous section, here in the non flat universe, we consider
the third form of interaction between dark matter and dark energy
$Q\propto H(\rho_{m}+\rho_{\Lambda})$. Using Eqs. (\ref{frid1}) and
(\ref{Freq2}), this form of interaction in non flat universe can be
written as $Q=\Gamma \rho_{\Lambda}$, where $\Gamma=3b^2H
\frac{1+\Omega_k}{\Omega_{\Lambda}}$.
 Taking the time derivative of
both side of Friedmann equation (\ref{frid1}) and using
(\ref{Freq2}, \ref{contmt}, \ref{contdt}) as well as the relation
$p_{\Lambda}=w_{\Lambda}\rho_{\Lambda}$, one can obtain
\begin{equation}\label{hdot}
\frac{\dot{H}}{H^2}=\Omega_k-\frac{3}{2}[1+\Omega_k+w_{\Lambda}\Omega_{\Lambda}]
\end{equation}
where $\Omega_k$ is given by
\begin{equation}\label{omk}
\Omega_k=\frac{a\gamma(1-\Omega_{\Lambda})}{1-a\gamma}, ~~~~~~~~~~
\gamma=\frac{\Omega_{k0}}{\Omega_{m0}}
\end{equation}
Inserting the interaction term $Q$ in the right hand side of
continuity equation (\ref{contdt}) and using the relations
(\ref{statet}), (\ref{hdot}), the equation of state for interacting
ghost dark energy in the non flat universe can be obtained as
\begin{equation}\label{state2}
w_{\Lambda}=\frac{2}{2-\Omega_{\Lambda}}\Big[-1+\frac{1}{2}(1+\Omega_k)(1-\frac{2b^2}{\Omega_{\Lambda}})-\frac{\Omega_k}{3}\Big]
\end{equation}
In the limiting case of  flat universe (i.e., $\Omega_k=0$),
Eq.(\ref{state2}) reduces to (\ref{statet1}), as expected. Using
(\ref{hdot}), the deceleration parameter $q$ in non flat case can be
obtained as
\begin{equation}\label{dece1}
q=-1-\frac{\dot{H}}{H^2}=\frac{1}{2}(1+\Omega_k)+\frac{3}{2}w_{\Lambda}\Omega_{\Lambda}
\end{equation}
The evolution of dark energy density in non flat universe is
obtained by tacking the time derivative of (\ref{denergyt}) and
using the ghost dark energy density (\ref{statet})
\begin{equation}\label{state99}
\dot{\Omega_{\Lambda}}=-\frac{\alpha\dot{H}}{3M_p^2H^2}
\end{equation}
Using (\ref{dece1}) and
$\dot{\Omega_{\Lambda}}=H\Omega_{\Lambda}^{\prime}$ results
\begin{equation}
\Omega_{\Lambda}^{\prime}=\Omega_{\Lambda}(1+q)
\end{equation}
where $q$ is defined in (\ref{dece1}). Tacking the time derivative
of Eq. (\ref{hdot}) and using (\ref{denergyt}), (\ref{contdt}),
(\ref{omk}) and (\ref{state2}) results
\begin{eqnarray}\label{ddoth}
\frac{\ddot{H}}{H^3}=\frac{9}{4}w_{\Lambda}\Omega_{\Lambda}(w_{\Lambda}\Omega_{\Lambda}+3)-
\frac{3}{2}\Omega_{\Lambda}(w_{\Lambda}^{\prime}-\Omega_kw_{\Lambda}\frac{\Omega_{\Lambda}-3/2}{\Omega_{\Lambda}-1})
+\Omega_k\frac{\Omega_{\Lambda}(\Omega_k+7)-10}{4(\Omega_{\Lambda}-1)}+\frac{18}{4}
\end{eqnarray}
Inserting Eqs. (\ref{dece1}) and (\ref{ddoth}) in Eq.
(\ref{stateft}), the former statefinder parameter $r$ for
interacting ghost dark energy in the non flat universe is obtained
as
\begin{equation}\label{stater1}
r=1+\frac{9}{4}w_{\Lambda}\Omega_{\Lambda}(w_{\Lambda}\Omega_{\Lambda}+1)-
\frac{3}{2}\Omega_{\Lambda}(w_{\Lambda}^{\prime}-\Omega_kw_{\Lambda}\frac{\Omega_{\Lambda}-3/2}{\Omega_{\Lambda}-1})
+\Omega_k\frac{\Omega_{\Lambda}(1+\Omega_k)-4}{4(\Omega_{\Lambda}-1)}
\end{equation}
Following \cite{evans}, we consider the parameter $s$ in the non
flat universe as follows
\begin{equation}\label{s5}
s=\frac{r-\Omega_t}{3(q-\Omega_t/2)}
\end{equation}
where $\Omega_t=1+\Omega_k$ is a total energy density as defined in
Friedmann equation. Obviously, in the limiting case of flat
universe, i.e., $\Omega_k=0$, the above definition is reduced to
(\ref{state1}).\\
Substituting Eqs. (\ref{dece1}) and (\ref{stater1}) in (\ref{s5})
gets
\begin{equation}\label{s6}
s=\frac{1}{2}(1+w_{\Lambda}\Omega_{\Lambda})-\frac{w_{\Lambda}^{\prime}}{3w_{\Lambda}}+
\frac{\Omega_k}{3(\Omega_{\Lambda}-1)}\left(\Omega_{\Lambda}-3/2+\frac{\Omega_k-3}{6w_{\Lambda}}\right)
\end{equation}
In the limiting case of flat universe, the above equations for the
statefinder parameter $\{s,r\}$ are reduced to those obtained in
previous section. Here in this section, we focus on the contribution
of spatial curvature on the evolution of ghost dark energy in the
$s-r$ and $q-r$ planes. For this aim we need to solve numerically
the relations (\ref{dece1}, \ref{stater1} and \ref{s6}). Note that
in these equations the dynamics of EoS parameter $w_{\Lambda}$,
density parameter $\Omega_{\Lambda}$ and spatial curvature parameter
$\Omega_k$ are given by (\ref{state2}), (\ref{denergyt}) and
(\ref{omk}), respectively.\\
In Fig.(3), we plot the statefinder diagram for different
contribution of spatial curvatures. The selected curves are plotted
by fixing $\Omega_{m0}=0.30$, $\Omega_{\Lambda 0}=0.70$ and varying
$\Omega_{k0}=0.02$, $\Omega_{k0}=0.00$ and $\Omega_{k0}=-0.02$
corresponding to the closed, flat and open universe, respectively. A
closed universe with a small positive curvature ( $\Omega_k=0.02$)
is compatible with some observations \cite{ben}. Here we ignore the
interaction between dark matter and dark energy and focus only on
the effect of contribution of spatial curvature on the evolution of
trajectories in statefinder plane. By expanding the universe, the
trajectories evolve from right to left. The parameter $r$ decreases,
then increases and reaches to the constant value $r=1$ at the late
time. The parameter $s$ decreases forever. The different
contributions of spatial curvature exhibit the different features in
the $s-r$ plane. The colored circles on the curves are the today's
value of $\{s_0,r_0\}$ for different spatial curvatures. One can see
that the today's value $\{s_0,r_0\}$ of interacting ghost dark
energy with different spatial curvatures is discriminated. We can
clearly identify the distance from a given dark energy model to the
standard flat-$\Lambda$CDM model by using the r(s) evolution
diagram. Fig.(3) shows that in the closed universe the distance of
the  present value $\{s_0,r_0\}$ from the location of $\Lambda$CDM
fixed point $\{s=0,r=1\}$ is shorter compare with other spatial
curvatures. The holographic dark energy model from the viewpoint of
statefinder diagnostic analysis in the non flat universe has already
been investigated in \cite{setar}.  By comparing Fig.(3) with upper
panel of Fig.(1) of \cite{setar}, we see the similarity of
evolutionary trajectories between ghost dark energy model and the
holographic model of dark energy (with the model parameter $c=1$) in
non flat universe.\\
Fig.(4) shows the evolutionary trajectories of interacting ghost
dark energy in $q-r$ plane for different contributions of spatial
curvature of the universe. By expanding the universe the
trajectories evolve from right to left, the parameter $r$ first
decreases then increases and the parameter $q$ decreases from the
positive value at the early time (indicating the decelerated phase
of expansion) to the negative value at the at late time ( denoting
the accelerated phase). In $q-r$ plane, the interacting ghost model
is discriminated for different contribution of spatial curvatures.
The current value of statefinder pair $\{q_0, r_0\}$ is also
distinguished in different spatial curvatures of the universe. The
value of $\{q_0, r_0\}$ is larger in closed universe
($\Omega_k=0.02$) compare with flat ($\Omega_k=0.00$) and open
($\Omega_k=-0.02$) universe.

\begin{center}
\begin{figure}[!htbp]
\includegraphics[width=10cm]{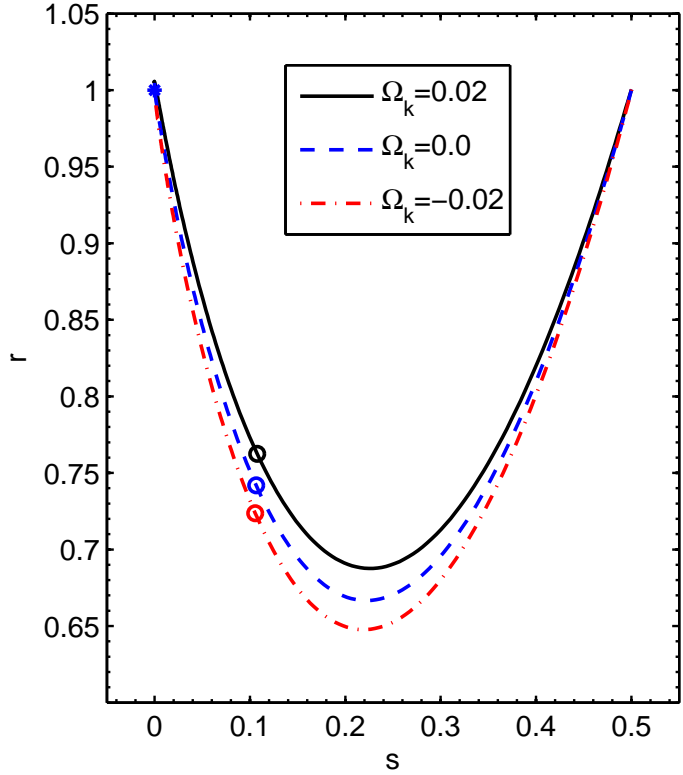}\\
~~~~~~ ~~~~~~\caption{The evolutionary trajectories of ghost dark
energy model in $s-r$ plane for different contributions of spatial
curvatures $\Omega_{k0}=0.02$ (closed universe), $\Omega_{k0}=0.00$
(flat universe), $\Omega_{k0}=-0.02$ (open universe). Here we set
$\Omega_{m0}=0.3$ and $\Omega_{\Lambda 0}=0.7$. The colored circle
points are the location of present value $\{s_0, r_0\}$ for
different spatial curvature as indicated in legend. The location of
$\Lambda$CDM fixed point has been shown by star symbol.}
 \end{figure}
 \end{center}

\begin{center}
\begin{figure}[!htbp]
\includegraphics[width=10cm]{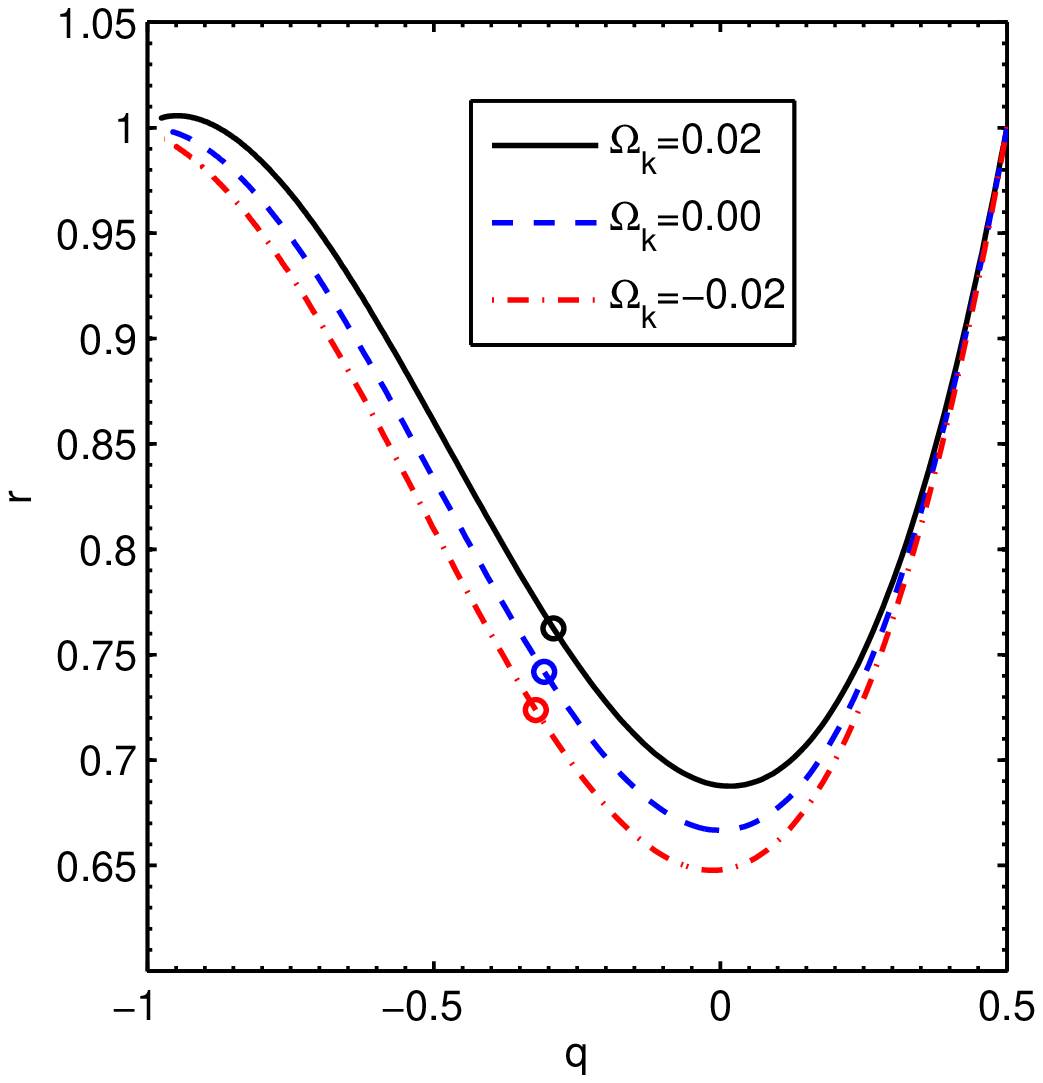}\\
~~~~~~ ~~~~~~\caption{ The evolutionary trajectories of ghost dark
energy model in $q-r$ plane for different contributions of spatial
curvatures $\Omega_{k0}=0.02$ (closed universe), $\Omega_{k0}=0.00$
(flat universe), $\Omega_{k0}=-0.02$ (open universe). Here we set
$\Omega_{m0}=0.3$ and $\Omega_{\Lambda 0}=0.7$. The colored circle
points are the location of present value $\{q_0, r_0\}$ for
different spatial curvature as indicated in legend.}
 \end{figure}
 \end{center}

\section{Interacting ghost model and observational constrains}
It is clear that constraining the parameterized model against the
observational data is model dependent. Hence some doubts usually
remain on the validity of the constraints on the derived quantities
such as the present day values of the deceleration parameter and the
age of the universe. In order to solve this problem, we use the
cosmography, i.e. the expansion of scale factor in Taylor series
with respect to the cosmic time. For this aim, the following
functions
\begin{eqnarray}\label{par}
  H &=& \frac{1}{a}\frac{da}{dt} \\
  q &=& -\frac{1}{a}\frac{d^2a}{dt^2}H^{-2} \\
  r &=& \frac{1}{a}\frac{d^3a}{dt^3}H^{-3} \\
  k &=& \frac{1}{a}\frac{d^4a}{dt^4}H^{-4} \\
  l &=& \frac{1}{a}\frac{d^5a}{dt^5}H^{-5}
\end{eqnarray}
which are namely the Hubble, deceleration, jerk, snap and lerk
parameters, respectively are introduced. The present values of the
above parameters can be used to describe the evolution of the
universe. For example, $q_0<0$ indicates the current accelerated
expansion of the universe and also $r_0$ allows to discriminate
between different dark energy models. Using the Union2 SNeIa data
set \cite{aman10} and the BAO data from the analysis of the SDSS
seventh release \cite{perc10} adding a prior on h from the recent
determination of the Hubble constant by the SHOES team \cite{ris09}
and the age of passively evolving galaxies \cite{gaz09}, the present
values of the above cosmographic parameters are constrain
observationally by using the Markov Chain Monte Carlo method
\cite{capa011}. The best fit values of the cosmographic parameters
are: \{$h=0.718$, $q_0=-0.64$, $r_0=1.02$,
$k_0=-0.39$, $l_0=4.05$\} (see table I of \cite{capa011} for more details).\\
Inserting the present values of $q_0=-0.64$ and $r_0=1.02$ in Eq.
(\ref{state1}), the present value of statefinder parameter $s$ is
obtained as $s_0=-0.006$. Therefore, observationally, the best fit
value of the current statefinder pair
is $\{s_0=-0.006,r_0=1.02\}$. In this section we compare the present
value of statefinder parameters \{s,r\} of interacting ghost dark energy that
has been constrained observationally in \cite{shaikh9} with
the above best fit value of current statefinder pair.\\
For this aim we use the best fit constrained values of the
cosmological parameters $\Omega_{m0}=0.35$, $\Omega_{\Lambda
0}=0.75$ and $b^2=0.08$ in the ghost dark energy model that have
recently been obtained in \cite{shaikh9} by using the data of
Supernova type Ia (SNIa) Gold sample, shift parameter of Cosmic
Microwave Background radiation (CMB) and the Baryonic Acoustic
Oscillation (BAO) peak from Sloan Digital Sky Survey (SDSS). In
Fig.(5) the evolutionary trajectories of interacting ghost dark
energy in $s-r$ plane (upper panel) and in $q-r$ plane (lower panel)
are plotted for the above best fit values of cosmological parameters
$\Omega_{m0}$, $\Omega_{\Lambda 0}$ and $b^2$. In $s-r$ diagram, the
evolutionary trajectory starts from $\{s =0.86, r = 0.67\}$ at the
past time, reaches to the $\{s =0.08, r = 0.74\}$ at the present
time (circle point) and ended at $\{s =0, r =1\}$ at the future. The
best fit observational value $\{s_0=-0.006,r_0=1.02\}$ in flat
universe is indicated by red-star symbol in this diagram. In $q-r$
diagram, the evolutionary trajectory starts from \{q = 0.4, r = 1\}
at the past ( corresponds to the decelerated expansion of the
universe), reaches to $\{q = -0.6, r =0.74\}$ at the present time
and ended at $\{q =-1, r=1\}$ at the late time ( corresponds to the
accelerated expansion). The best fit observational value
$\{q_0=-0.64,r_0=1.02\}$
is also indicated by red-star symbol in this diagram.\\
Now we compare the present value $\{s_0,r_0\}$ of constrained
interacting ghost dark energy model with other models of dark energy
which have been constrained and discussed from the viewpoint of
statefinder diagnostic. This comparison includes the interacting
ghost model, holographic, new holographic and generalized chaplygin
gas models of dark energy. These models have been constrained by
astronomical data of SNe+CMB+BAO+OHD experiments and also have been
discussed in $s-r$ diagram based on the constrained values of
cosmological and model parameters. This comparison also includes the
standard $\Lambda$CDM model as well as the best fit observational
value $\{s_0=-0.006,r_0=1.02\}$ in flat universe. The holographic
dark energy model with the constrained values ($c=0.84$,
$\Omega_{m0}=0.29$, $\Omega_{k0}=0.02$, where $c$ is the model
parameter of holographic dark energy) obtains the today's
statefinder pair as $\{s_0=-0.102, r_0=1.357\}$ \cite{setar}. The
new holographic dark energy model with the constrained values
($\Omega_{b}h^2=0.0228$, $\Omega_{m0}=0.2762$, $\Omega_{k0}=0.0305$,
$\Omega_{\Lambda 0}=0.6934$, $\alpha= 0.8824$, $\beta=0.5016$, where
$\alpha$ and $\beta$ are the parameters of model) results the
today's statefinder pair as $\{s_0=-0.13, r_0=1.46\}$
\cite{malek20}. The generalized chaplygin gas dark energy ( GCG
model) in the flat universe with the constrained values ($A_s =
0.76$, $\alpha=0.033$, $\Omega_{b}h^2= 0.0233$, $H_0 = 69.97$, where
$A_s$ and $\alpha$ are the parameters of the model) gives the
today's statefinder pair as $\{s_0=-0.007, r_0=1.026\}$
\cite{malek11}. Note that the GCG model is constrained in the flat
universe, but other models are constrained in general non-flat
universe. Fig.(6) shows the location of the present statefinder pair
$\{s_0,r_0\}$ for the above constrained models as indicated in
legend. The standard $\Lambda$CDM model and also the best fit
observational value $\{s_0=-0.006,r_0=1.02\}$ in flat universe are
indicated by black and red star symbols, respectively. One can
conclude that the $\Lambda$CDM model $\{s=0,r=1\}$ has a shortest
distance to the best fit observational value
$\{s_0=-0.006,r_0=1.02\}$ compare to other dynamical dark energy
models. Also, the interacting ghost dark energy model has a shorter
distance from $\{s_0=-0.006,r_0=1.02\}$ compare with the holographic
and new holographic dark energy models. Among the dynamical dark
energy model, the GCG model has a shortest distance from the
location of observational value in $s-r$ plane.

\begin{center}
\begin{figure}[!htbp]
\includegraphics[width=10cm]{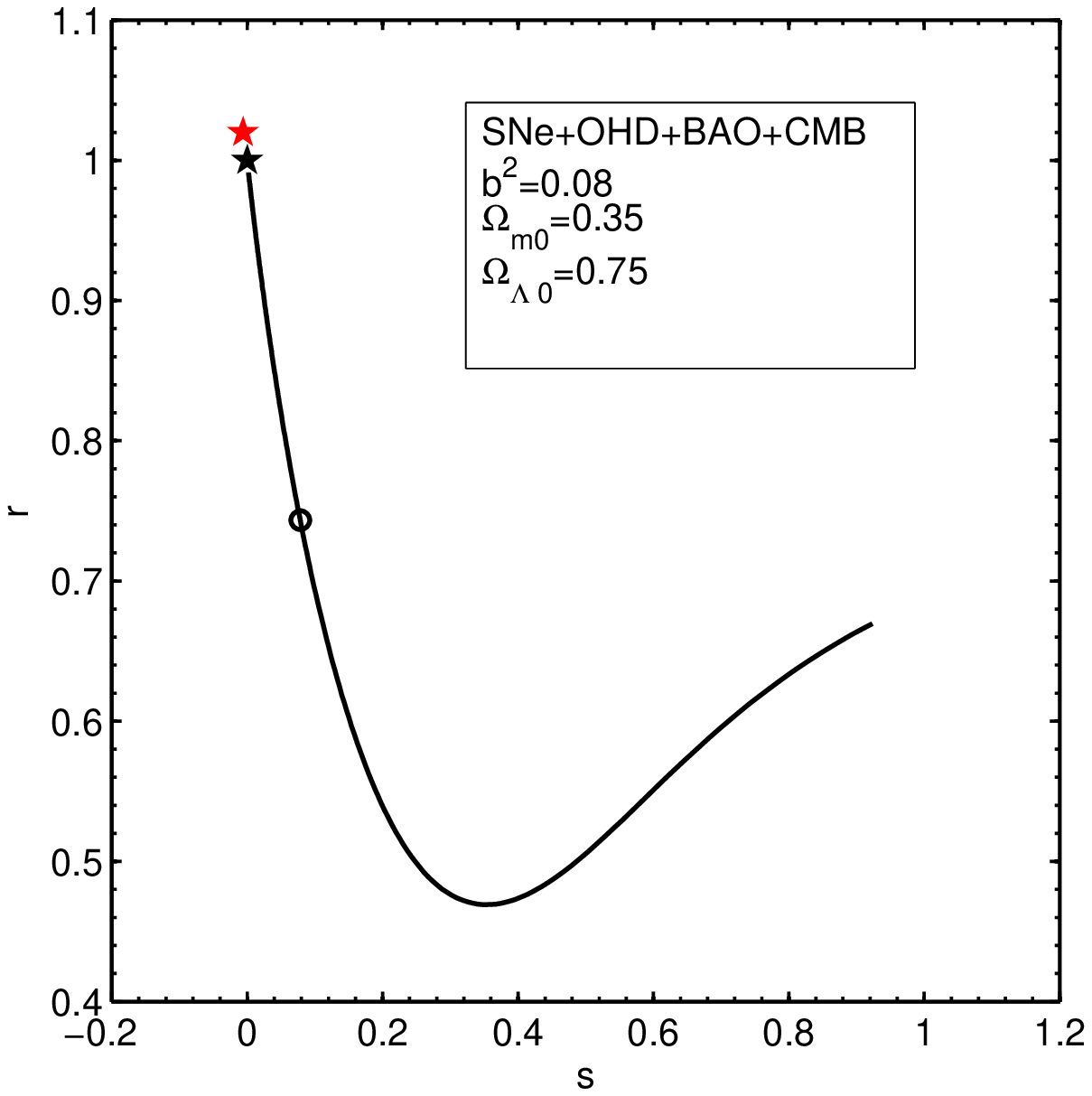}
\includegraphics[width=10cm]{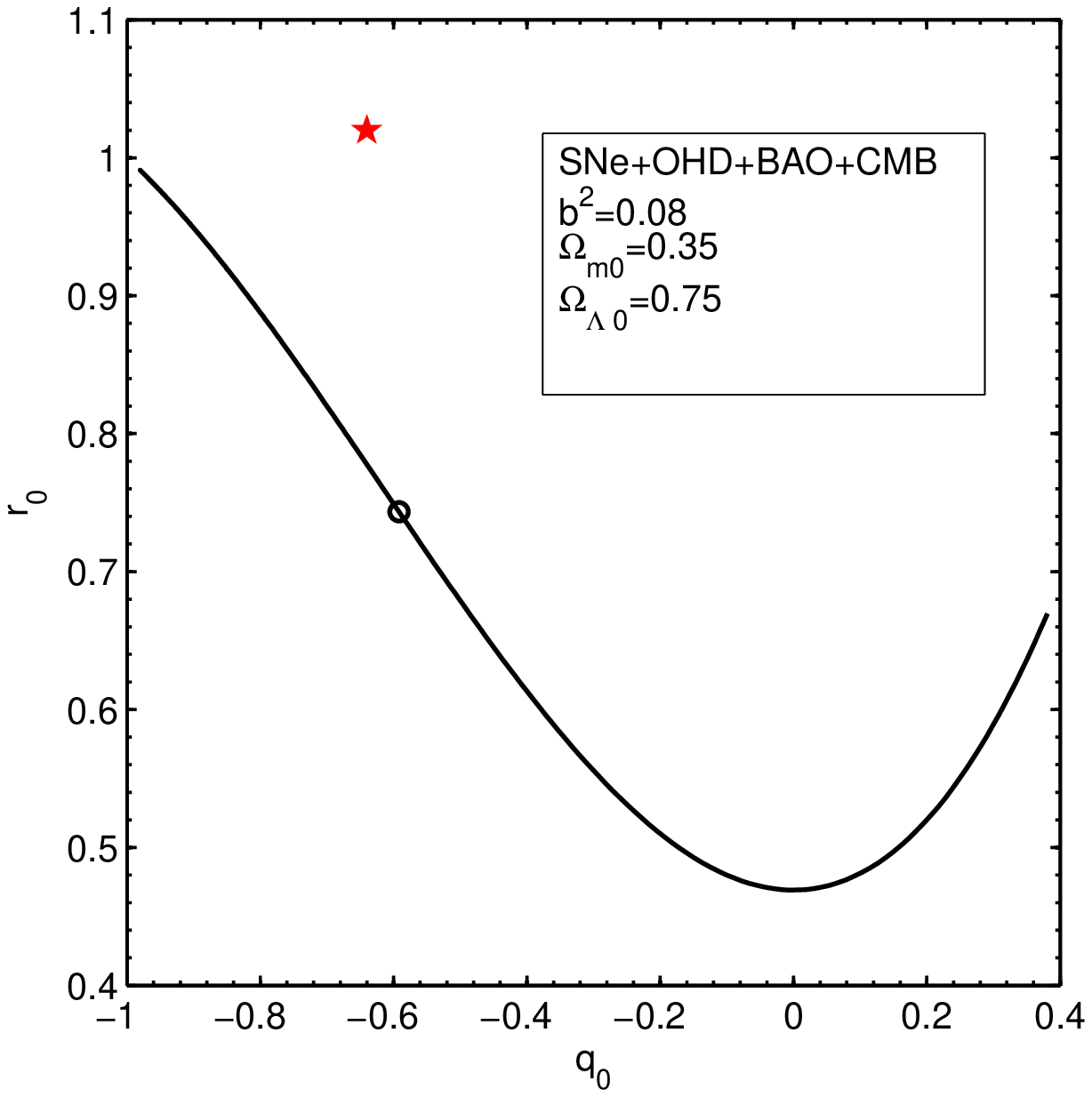}\\
~~~~~~ ~~~~~~\caption{The statefinder diagrams $r(s)$ (upper panel)
and $r(q)$ (lower panel) for interacting ghost dark energy model.
The evolutionary trajectories are plotted in the light of best fit
result of SNe + OHD + BAO + CMB, $\Omega_{\Lambda 0}=0.75$,
$\Omega_{m0}=0.35$ and $b^2=0.08$ . The circle points on the curves
show the today's value $\{s_0, r_0\}$, upper panel, and $\{q_0,
r_0\}$, lower panel. For comparison, the standard $\Lambda$CDM model
has been shown by black-star symbol and the constrained
observational value of $\{s_0, r_0\}$ and $\{q_0, r_0\}$ are
indicated by red-star symbol in these diagrams}
 \end{figure}
 \end{center}

\begin{center}
\begin{figure}[!htbp]
\includegraphics[width=10cm]{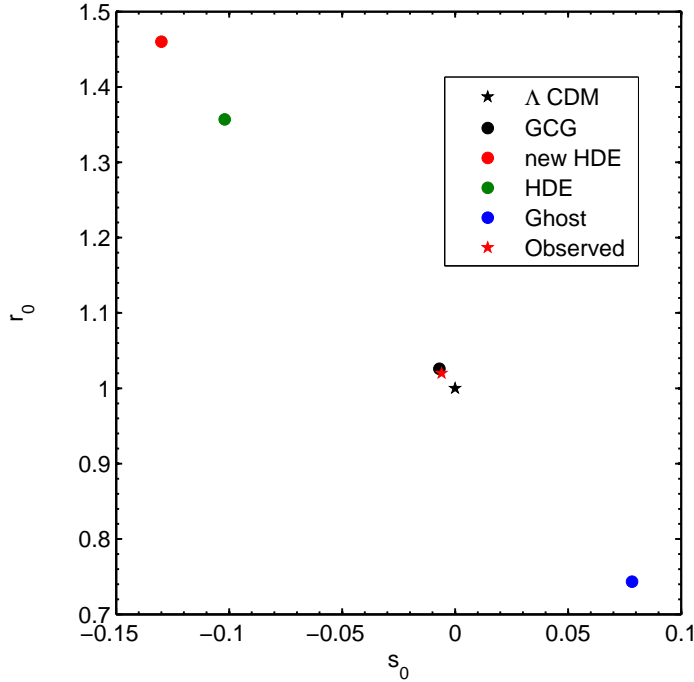}\\
~~~~~~ ~~~~~~\caption{The present value of $\{s_0, r_0\}$  in the
light of best fit result of SNe + OHD + BAO + CMB observations for
different dark energy model as indicated in legend. The location of
standard $\Lambda$CDM model and constrained observational value
$\{s_0, r_0\}$ have been shown by black and red star symbols,
respectively.}
 \end{figure}
 \end{center}
\newpage
\newpage
\section{conclusion}
Summarizing this work, we investigated the interacting ghost dark
energy model in statefinder $s-r$ and $q-r$ diagrams. The
statefinder analysis can discriminate the interacting ghost dark
energy model for different values of interaction parameter as well
as the different spatial curvatures of the universe. Like
holographic \cite{zhang056,zhang057,setar}, new agegraphic
\cite{zhang058,malek12}, chaplygin gas \cite{chap2}, generalized
chaplygin gas \cite{malek11} and yang-mils \cite{yang} models of
dark energy, the ghost dark energy model mimics the standard
$\Lambda$CDM model at the late time. The evolution of ghost dark
energy model in $s-r$ plane is similar to holographic model of dark
energy with the model parameter $c=1$. Different values of
interaction parameter obtains the different evolutionary
trajectories in $s-r$ and $q-r$ planes. The evolutionary
trajectories $r(s)$ and $r(q)$ for interacting ghost dark energy
model in different closed, flat and open universe has also been
investigated. We have shown that different contribution of spatial
curvatures give the different evolutionary trajectories in $s-r$ and
$q-r$. The spatial curvature can also influence the present value of
statefinder parameters $\{s_0,r_0\}$ and $\{q_0,r_0\}$ in these
planes. Eventually, we performed the statefinder diagnostic for the
interacting ghost model constrained by observational data. We
conclude that the $\Lambda$CDM model $\{s=0,r=1\}$ has a shortest
distance to the best fit observational value
$\{s_0=-0.006,r_0=1.02\}$ compare with other dynamical dark energy
models. Therefore the models of dark energy whose curerent
statefinder values locate far from the $\Lambda$CDM point can be
ruled out. The interacting ghost dark energy model has a shorter
distance from $\{s_0=-0.006,r_0=1.02\}$ compare with the holographic
and new holographic dark energy models. Among the above dynamical
dark energy models, the GCG model has a shortest distance from the
location of observational statefinder pair (i.e.,
$\{s_0=-0.006,r_0=1.02\}$).

\end{document}